\documentclass{article}

\usepackage[preprint]{neurips_2019}

\usepackage[utf8]{inputenc} % allow utf-8 input
\usepackage[T1]{fontenc}    % use 8-bit T1 fonts
\usepackage{hyperref}       % hyperlinks

\usepackage{booktabs}       % professional-quality tables
\usepackage{amsfonts}       % blackboard math symbols
\usepackage{nicefrac}       % compact symbols for 1/2, etc.
\usepackage{microtype}      % microtypography

\usepackage{subcaption} % for subfigure
\usepackage{hyperref}
\usepackage{graphicx}
\usepackage{amsmath}
\usepackage{etoolbox,refcount}
\usepackage{multicol}

\usepackage{url}            % simple URL typesetting
\usepackage{relsize}
\usepackage{tikz}

\usepackage{amssymb}% http://ctan.org/pkg/amssymb
\usepackage{pifont}% http://ctan.org/pkg/pifont
\newcommand{\cmark}{\ding{51}}%
\newcommand{\xmark}{\ding{55}}%

\newcommand{\norm}[1]{\left\lVert#1\right\rVert}

\newcounter{countitems}
\newcounter{nextitemizecount}
\newcommand{\setupcountitems}{%
  \stepcounter{nextitemizecount}%
  \setcounter{countitems}{0}%
  \preto\item{\stepcounter{countitems}}%
}
\makeatletter
\newcommand{\computecountitems}{%
  \edef\@currentlabel{\number\c@countitems}%
  \label{countitems@\number\numexpr\value{nextitemizecount}-1\relax}%
}
\newcommand{\nextitemizecount}{%
  \getrefnumber{countitems@\number\c@nextitemizecount}%
}
\newcommand{\previtemizecount}{%
  \getrefnumber{countitems@\number\numexpr\value{nextitemizecount}-1\relax}%
}
\makeatother

{\end{itemize}%
\unskip\computecountitems\ifnumcomp{\previtemizecount}{>}{3}{\end{multicols}}{}}

{\end{itemize}%
\unskip\computecountitems\ifnumcomp{\previtemizecount}{>}{2}{\end{multicols}}{}}

\title{Auditory Separation of a Conversation from Background via Attentional Gating}

\author{%
  Shariq Mobin\\
  Redwood Center for Theoretical Neuroscience\\
  Helen Wills Neuroscience Institute\\
  University of California, Berkeley\\
  Berkeley, CA 94720\\
  \texttt{shariq.mobin@gmail.com} \\
  \And
  Bruno Olshuasen \\
Redwood Center for Theoretical Neuroscience\\
  University of California, Berkeley\\
  Berkeley, CA 94720\\
}

\begin{document}

\maketitle

\begin{abstract}
We present a model for separating a set of voices out of a sound mixture containing an unknown number of sources. Our Attentional Gating Network (AGN) uses a variable attentional context to specify which speakers in the mixture are of interest. The attentional context is specified by an embedding vector which modifies the processing of a neural network through an additive bias. Individual speaker embeddings are learned to separate a single speaker while superpositions of the individual speaker embeddings are used to separate sets of speakers. We first evaluate AGN on a traditional single speaker separation task and show an improvement of 9\% with respect to comparable models. Then, we introduce a new task to separate an arbitrary subset of voices from a mixture of an unknown-sized set of voices, inspired by the human ability to separate a conversation of interest from background chatter at a cafeteria. We show that AGN is the only model capable of solving this task, performing only 7\% worse than on the single speaker separation task.
\end{abstract}

\section{Introduction}
Single-channel audio source separation is a challenging speech processing task, where a system is required to separate the speaker signals from one another given only the mixture of the signals. The separation performance has increased significantly in the last few years thanks to advances in deep learning models such as deep clustering \citep{he2016deep}, deep attractor network \citep{chen2017deep}, and permutation invariant training \citep{yu2017permutation}. This line of research has largely focused on separating a mixture of only two or three voices talking over one another simultaneously. In these models the number of speakers in the mixture must be known a priori. However, in more realistic scenarios there are many more voices and the number of voices is not known before hand, as in a noisy cafeteria \footnote{See supplementary audio or \url{https://soundcloud.com/anon-ymous-647326941/conversation-over-cafeteria}}. In the vocabulary of \cite{bregman1994auditory} the number of sources in this example is ill-defined: at some time points the voices in the background are grouped into background chatter and at other time points voices emerge as their own source separate from the background chatter. This renders deep learning models that assume the number of sources is known impractical.

\begin{figure}[h]
    \centering % <-- added
    \begin{subfigure}{.4\textwidth}
      \includegraphics[scale=0.5]{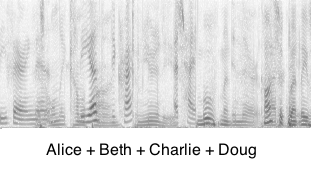}
    \end{subfigure}\hfil % <-- added
    \centering
    \begin{subfigure}{.4\textwidth}
      \includegraphics[scale=0.5]{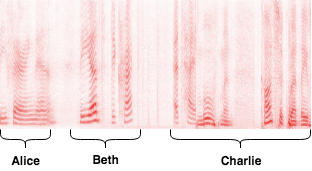}
    \end{subfigure}\hfil % <-- added
\caption{Example input and output of the speaker-set attention model. Time is on the x-axis and Frequency is on the y-axis.  \emph{Left}: Input Spectrogram of the mixture which contains four voices. A second embedding vector input specifies that the voices of Alice, Beth, and Charlie should be separated (not shown). \emph{Right}: Output of AGN, an estimation of the spectrogram for the three voices of interest. The model is unaware of the total number of speakers in the mixture.}
\label{fg:attention_example}
\end{figure}

Humans solve the separation problem everyday as they attend to the voices of their friends and family over the voices of others in noisy environments. It is thought that attention is the core idea that makes this possible for humans \citep{fritz2007auditory}. The idea is to decide which voices are of interest and only separate those, bypassing the need to know the total number of sources. The mammalian brain is hypothesized to utilize top-down feedback connections which carry a neural signal representing this decision to modify bottom-up processing to separation the voices of interest \citep{oatman1971role}. Here we develop a computational model of this process called the Attentional Gating Network or AGN. In AGN the top-down neural signal comes in the form of an embedding vector which introduces a variable additive gating factor to the bottom-up pathway. See Figure \ref{fg:attention_example} for an example of the input and output of AGN.

Deep learning based speech separation systems are often classified into two groups: speaker-independent and speaker-biased speech separation.

\paragraph{Speaker-Independent} In this approach the model attempts to separate the sound mixture into all of it's constituent sources at once. Models that follow this approach project time-frequency activations of the mixture spectrogram into a high dimensional space where K-means can be applied to extract the K sources in the mixture. Using deep learning, these methods have been shown to be extremely effective, e.g. \citep{he2016deep, chen2017deep}. However, a major drawback of this approach is that the number of sources in the mixture, $K$, must be known before hand otherwise K-means cannot be applied. It has not been shown how $K$ could be estimated online in real-time. Doing so would require estimating $K$ at each time point and matching clusters across time which seems nontrivial at best.

\paragraph{Speaker-Biased} By contrast, the speaker-biased approaches only extract a target speaker's voice, while all other voices are considered as interference. In order to specify the target some form of prior knowledge of the target is needed. One form of prior knowledge is a segment of reference audio from the target speaker. The advantage of the speaker-biased method is that it is agnostic to the number of inputs in the mixture. Using deep learning, this approach performs very well, comparable to speaker-independent models as shown in the Speech Extraction Network \citep{xiao2019single}, VoiceFilter \citep{wang2018voicefilter}, and SpeakerBeam \citep{delcroix2018single} models.

The reference audio method is only one approach to specify the target. Common alternatives are to use an additional input feature as in the i-vectors approach of \cite{saon2013speaker, senior2014improving} or to include additional parameters per speaker \cite{ochiai2014speaker, gemello2007linear}. AGN uses a combination of both of these, similar to \cite{vincent2014blind}, by allocating an embedding vector per speaker that is learned to specify the target output.

Since reference audio is not used, enabling AGN to operate on new speakers is not obvious. To generalize from the closed speaker-set to the open speaker-set ideas from the fine-tuning and meta-learning literature are used to learn to adapt only the parameters necessary for separation. In particular, the embedding vector can be viewed as augmented memory \cite{santoro2016meta} that needs to be trained for each new task, or in this case, new speaker. Of future interest would be the meta-learning approach of \cite{finn2017model} and the matching network approach of  \cite{vinyals2016matching}.

We make the following contributions:

\textbf{Speaker-Set Separation }\hspace{4mm} Extend current speaker-biased approaches to handle the simultaneous separation of an arbitrary number of speakers rather than just one, using embedding superpositions. This allows speaker-biased approaches to separate many speakers at once like speaker-independent methods.

\textbf{Latent Speaker Embeddings}\hspace{4mm} Show how embeddings can be learned through inference in a deep learning context. This is in contrast to the reference audio approach which relies on a recurrent neural network to compute an embedding. This method results in 9\% improvement on the speaker-biased task.

\textbf{Flexibility to Input}\hspace{4mm} Introduce a new stochastic speaker-set-biased task which demonstrates how speaker-biased models are agnostic to the number of sources in the input mixture. This also demonstrates that attention can be about more than computational efficiency - Models based off of attention can perform computations not possible by models that forego it.

\section{Speaker-Set Separation}
\subsection{Stochastic Speaker-Set-Biased Task Definition}
In \cite{xiao2019single} the task is to extract the target speaker's voice from a mixed speech waveform that contains exactly two voices. Here, we extend this task to extract a set of target speakers' voices from a mixed speech waveform that contains an unknown number of sources. The characteristics of the speakers are known through their profiles, which each contain a set of utterances collected from a dataset where the speaker's label is available. We investigate two scenarios with respect to the utterances:

\textbf{Well-Known Speakers} - Substantial utterances on these speakers, $\sim20$ minutes of data per speaker.

\textbf{New Speakers} - Minimal utterances on these speakers, $100$ seconds of data per speaker.

In this task, we can think of training on the \textit{well-known} speakers as pre-training and training on the \textit{new} speakers as fine-tuning. The idea is that after pre-training we want to be able to quickly and data-efficiently learn to new tasks, i.e. separate \textit{new} speakers.

\subsection{Speaker-Set Extraction Framework}
\subsubsection{Setup}
\textbf{Notation:} For a tensor $T \in \mathbf{R}^{A \times B \times C}$: $T_{\cdot, \cdot, c} \in \mathbf{R}^{A \times B}$ is a matrix, $T_{a,\cdot, c} \in \mathbf{R}^B$ is a vector, and $T_{a,b,c} \in \mathbf{R}$ is a scalar.

Let $N$ be the number of speakers. Each mixture input signal is the sum of a target and interference signal $x = t + d$, $x \in \mathbb{R}^{\tau}$. The target and interference signals themselves are the sum of $G$ and $H$ waveforms respectively, each waveform coming from a different speaker. Note $G \ll N, H \ll N$. A special $G$-hot vector, $B \in \{0,1\}^N$, is used to specify which speakers should be output of the model. More formally:
\begin{eqnarray*}
z_k &\bar{\sim}& \mathcal{U}\{0, N-1\}, \hspace{4mm} k \in [0, G+H) \\
t &=& \sum_{k=0}^{G-1} s_k, \hspace{9mm} s_k \sim S_{z_k} \\
%\end{eqnarray*}
%\begin{eqnarray*}
d &=& \sum_{k=G}^{G+H -1} s_k, \hspace{4mm} s_k \sim S_{z_k} \\
x &=& t + d \\
B[z_k] &=&
    \begin{cases}
        1, \hspace{4mm} \text{if } k < G\\
        0, \hspace{4mm} \text{else} \\
    \end{cases}
\end{eqnarray*}
where $\bar{\sim}$ denotes sampling without replacement, $\mathcal{U}\{\}$ denotes the discrete uniform distribution, and $S_i$ corresponds to another discrete uniform distribution over the waveforms of speaker $i$. $\mathbb{X} \in \mathbf{R}^{F \times T} $ is the spectrogram of $x$, computed using the Short-time Fourier transform (STFT), where $F$ corresponds to frequencies and $T$ to time. Analogously $\mathbb{T} \in \mathbf{R}^{F \times T}$ is the spectrogram of $t$. To formulate the speaker-biased task of previous work we simply set $G=1$, and $H=1$. To define our new stochastic speaker-set-biased task we instead sample $G$ and $H$ from a distribution for each example:
\begin{eqnarray}
G &\sim& \mathcal{U}\{1, 3\},  \\
H &\sim& \mathcal{U}\{1, 3\} \label{eq:g_sample}
\end{eqnarray}

Before inputting $\mathbb{X}$ to the model we apply a compression function, $f_c$. Here we use a power-law  nonlinearity to compress, inspired by \cite{wilson2018exploring}:
\begin{eqnarray*}
f_c(u) &:=& u^p \\
\bar{\mathbb{X}} &=& f_c\Big(abs\big(STFT(x)\big)\Big), \hspace{8mm} \bar{\mathbb{T}} = f_c\Big(abs\big(STFT(t)\big)\Big)
\end{eqnarray*}

The power law is desirable because it has the following property: $\frac{f_c(\alpha x)}{f_c(x)} \hspace{2mm} \forall x \in \mathbb{R}$ with fixed $\alpha \in \mathbb{R}$. This makes quieter sounds as relevant as loud sounds, unlike the log function.

\subsubsection{Speaker-Set Embeddings}
All $N$ speakers have a $K$-dimensional speaker embedding, $E \in \mathbb{R}^{N,K}$. To specify a single target speaker, $G=1$, the desired speaker embedding is supplied as an additional input to the model. The embeddings are parameters that must be learned in addition to the neural network parameters.

\begin{figure*}[t]
    \centering
    \includegraphics[scale=0.35]{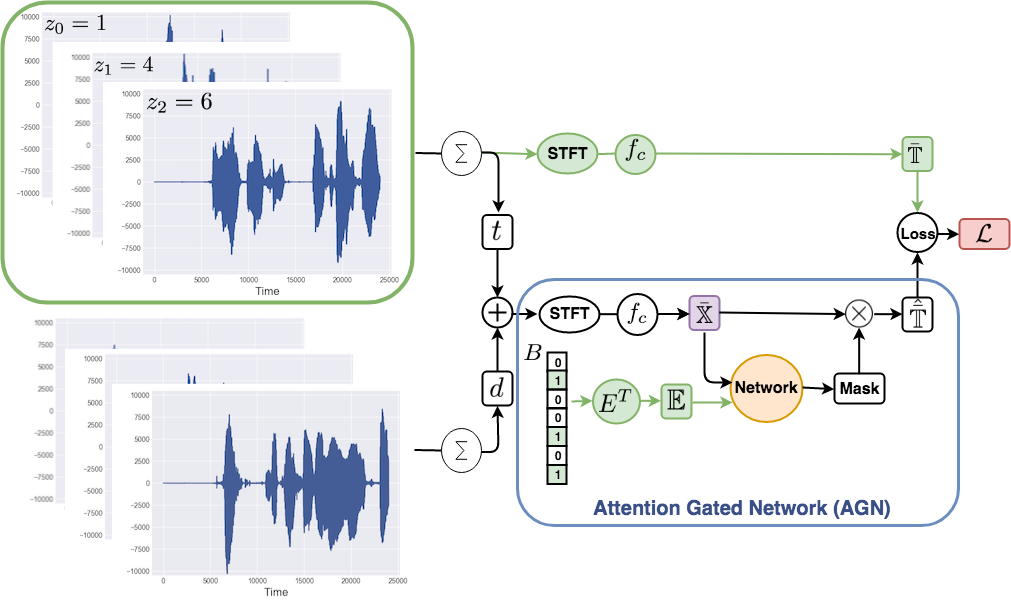}
    \caption{Training pipeline for the Attentional Gating Network (AGN). Circles denote operations and rounded squares represent variables. The components colored in green represent supervised information about the desired target speaker. In this diagram $G=3$, $H=3$, and $N=7$. The abs function is skipped in the diagram for brevity. $f_c$ corresponds to the compression function used. See text for details. }
    \label{attention_schematic}
\end{figure*}

To specify a set of target speakers, $G > 1$, a speaker-set embedding, $\mathbb{E} \in \mathbb{R}^K$, is computed using a superposition of the desired individual speaker embeddings, by utilizing $B$. Thus no additional parameters are necessary. The full model works as follows:
\begin{eqnarray*}
\mathbb{E} &=& E^T B \\
M_{f,t} &=& \sigma (f_{NN}(\bar{\mathbb{X}}, \mathbb{E}; \theta)) \\
\hat{\bar{\mathbb{T}}}_{f,t} &=& M_{f,t} \odot \bar{\mathbb{X}}_{f,t} \\
\mathcal{L} &=& \norm{\bar{\mathbb{T}} - \hat{\bar{\mathbb{T}}}}^2_F \label{loss}
\end{eqnarray*}

where superscript $T$ denotes transpose, $f_{NN}$ is a neural network, $\theta$ represents the parameters of the neural network, $M \in [0,1]^{F \times T}$ is a mask, $\odot$ denotes element-wise product, and $||\cdot||_F$ indicates the Frobenius norm. See Figure \ref{attention_schematic} for an overview of AGN. The target estimate, $\hat{t}$, can then be recovered using the inverse Short-Time Fourier Transform of $f_c^{-1}(\hat{\bar{\mathbb{T}}})$ with the mixture phase of $\mathbb{X}$.

\subsubsection{Pre-Training (Well-Known Speakers)}
During pre-training the neural network parameters and the \textit{well-known} speaker embedding parameters are trained jointly:
\begin{eqnarray*}
\Theta &:=& [\theta, E_{i, \cdot}, ..., E_{j, \cdot}], \hspace{2mm} i = 0, j = N - 1 \hspace{4mm} \text{// N is the number of \textit{well-known} speakers.} \\
\Theta &=& \Theta - \eta \frac{\partial L}{\partial \Theta}
\end{eqnarray*}

where $\eta$ is the learning rate.

\subsubsection{Fine-Tuning (New Speakers)}
During fine-tuning the parameters can be updated in one of two possible modes:

\paragraph{Conventional}
During conventional fine-tuning the network parameters are initialized to the converged pre-training parameters and a set of \textit{new} speaker embeddings are initialized randomly. All parameters are updated to optimize performance on only the \textit{new} speakers.

\paragraph{Robust} In the robust mode all parameters are initialized the same way as conventional fine-tuning. However, only the \textit{new} speaker embedding parameters are updated in this mode.
\begin{eqnarray*}
\Theta &:=& [E_{i, \cdot}, ..., E_{j, \cdot}], \hspace{2mm} i = 0, j = N - 1 \hspace{4mm} \text{// N is the number of \textit{new} speakers.} \\
\end{eqnarray*}
The robust mode allows for AGN to learn \textit{new} speakers, or tasks, without interfering with it's performance on the already-learned \textit{well-known} speakers. By only updating the new embedding parameters performance on old, \textit{well-known} speakers is preserved. This approach prevents the catastrophic forgetting of old tasks, as described in \cite{french1999catastrophic, mccloskey1989catastrophic}.

\begin{figure*}[t]
    \centering
    \includegraphics[scale=0.36 ]{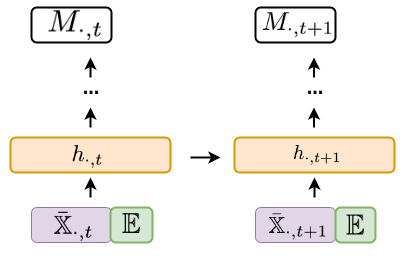}
    \caption{Illustration of how $\mathbb{E}$ is incorporated into a recurrent neural network. At each time step $\mathbb{E} = E^T B$ is appended to the frequency input, $\bar{\mathbb{X}}_{\cdot, t}$, in order to specify the $G$ targets marked in $B$.}
    \label{LSTM-Append-Model}
\end{figure*}

\subsection{Network Architecture}
We use a neural network, BLSTM-FC, which is composed of multiple layers of a Bi-Directional LSTM \citep{graves2005bidirectional} followed by multiple fully connected layers. The speaker-set embedding, $\mathbb{E}$, is appended to the input at each time point. This is similar to the approach taken in \cite{xiao2019single} and \cite{wang2018voicefilter}.
\begin{eqnarray*}
f_{NN}(\bar{\mathbb{X}}, \mathbb{E}; \theta) &=& \text{BLSTM-FC}(A) \\
A_{\cdot, t} &:=& [\bar{\mathbb{X}}_{\cdot, t}, \mathbb{E}], \hspace{3mm} A \in \mathbb{R}^{(F+N) \times T}
\end{eqnarray*}

For a recurrent neural network this amounts to an additional bias term which corresponds to a weight matrix, $W_{eh}$, multiplied by the speaker-set embedding:
\begin{eqnarray*}
h_{t+1} &=& tanh(W_{hh} h_{t} + W_{ih} \bar{\mathbb{X}}_{\cdot, t+1}  + b) \hspace{17mm} \text{// No Attention (normal RNN)} \\
h_{t+1} &=& tanh(W_{hh} h_{t} + W_{ih}\bar{\mathbb{X}}_{\cdot, t+1}  + b + W_{eh} \mathbb{E}) \hspace{4mm} \text{// Add Attention}
\end{eqnarray*}

where $W_{hh}$ is the hidden to hidden weight matrix, $W_{ih}$ is the input to hidden weight matrix, and $b$ is the normal bias term. Thus the embedding vector introduces an additive gain factor for the first recurrent layer neurons. By specifying different gain factors, AGN's computation vastly changes in order to extract an arbitrary subset of speakers from a mixture. See Figure \ref{LSTM-Append-Model} for an illustration.

\section{Experimental Results}
We evaluate the models on a single-channel speaker-biased and speaker-set-biased task. The mixture waveforms are transformed into a time-frequency representation using the Short-time Fourier Transform (STFT), after which the absolute value and power law function are applied, with $p=0.3$. The STFT is computed with 32ms window length, 16ms hop size, and the Hann window. The waveform is downsampled to 8kHz to reduce computational cost. The length of the waveforms is set to $\tau = 40000$. We use TensorFlow to compute the STFT and build the neural networks \citep{abadi2016tensorflow}.

Results are reported using the signal-to-distortion ratio (SDR) which we define as the scale-invariant signal-to-noise ratio (SI-SNR) as suggested by \cite{le2018sdr}, and is also used in [\cite{he2016deep}, \cite{chen2017deep}, \cite{luo2018tasnet}]:
\begin{eqnarray*}
t^{proj} &=& \frac{\langle \hat{t}_, t \rangle t}{\norm{t}^2} \\
e_{noise} &=& \hat{t} - t^{proj} \\
\textrm{SDR} := \textrm{SI-SNR} &:=& 10 \hspace{1mm} log_{10} \frac{\norm{t^{proj}}^2}{\norm{e_{noise}}^2}
\end{eqnarray*}

where $\hat{t}$ is the target estimate computed by AGN, and $t$ is the true target.

\subsection{Datasets}
To train and evaluate the models the LibriSpeech dataset of  \cite{panayotov2015librispeech} is used. The train-clean-360 and test-clean portions of the dataset are used for the experiments. The train-clean-360 data set contains 360 hours of clean speech data spread over $N = 916$ speakers, which we call the \textit{well-known} speakers. We call the $N = 40$ speakers in the test-clean dataset \textit{new} speakers. We split the \textit{well-known} and \textit{new} speakers into two portions each for training and evaluation:

\textbf{Well-Known Speakers, $N = 916$}
    \begin{itemize}
        \item \textit{well-known}-eval: The first 100 seconds of each speaker in train-clean-360
        \item \textit{well-known}-train: The remaining seconds of data ($\gg$ 100) of each speaker in train-clean-360
    \end{itemize}

\textbf{New Speakers, $N = 40$}
    \begin{itemize}
        \item \textit{new}-train: The first 100 seconds of each speaker in test-clean
        \item \textit{new}-eval: The remaining seconds of data ($\gg$ 100) of each speaker in test-clean
    \end{itemize}

During training the target and interferer signal are mixed with a signal-to-noise ratio (SNR) uniformly sampled from [-5dB, 5dB], following \cite{xiao2019single}. We use the BLSTM-FC neural network, with $5$ BLSTM layers, $3$ fully connected layers, and $512$ units for all layers. The embedding size, $K$, is also set to $512$.  RMSProp \citep{tieleman2012lecture} is used to optimize the network. We start with a learning rate of $3\mathrm{e}{-4}$ and decay exponentially every $3000$ steps at a decay rate of $0.95$.

\subsection{Experiments}
\subsubsection{Speaker-Biased Task - Single Speaker in Target and Interferer}
In these experiments there is a single speaker in the target and interferer, $G=1, H=1$, which is identical to the speaker-biased task of \cite{xiao2019single}.

\textbf{Pre-Training}
We begin by evaluating the model on the \textit{well-known} speaker set. The results are compared with the baseline speaker-biased model of \cite{xiao2019single} in Table \ref{single-result}. They indicate that AGN performs substantially better than other speaker-biased models. An audio sample is available \footnote{See supplementary audio or \url{https://soundcloud.com/anon-ymous-647326941/sets/single-attention-example}}.

\begin{table}
  \caption{Pre-Training results for the single-speaker target and interferer experiment. The SDR is reported for AGN and Xiao-Attention on the speaker-biased task. All datasets come from LibriSpeech and are comparable.}
  \label{single-result}
  \centering
  \begin{tabular}{lllll}
    \toprule
 %   \multicolumn{2}{c}{Part}                   \\
%    \cmidrule(r){1-2}
    & \textit{well-known}-eval & test-clean \\
    \midrule
    AGN             & 10.6 dB & x \\
    Xiao-Attention             & x & 9.8 dB \\
    \bottomrule
  \end{tabular}
\end{table}

\textbf{Fine-Tuning}
Next, the model from the previous experiment is fine-tuned  to separate \textit{new} speakers using conventional fine-tuning and robust fine-tuning. The results are shown in Table \ref{new-speaker-result}. We compare the results to the baseline attention model of \cite{xiao2019single}. Interestingly, the robust fine-tuning method which only optimizes the embedding parameters is able to perform as well as the conventional method that optimizes all parameters.

\begin{table}
  \caption{Fine-Tuning results for the single-speaker target and interferer experiment. The SDR for two types of fine-tuning are reported along with the baseline model of Xiao-Attention model on the speaker-biased task. All datasets come from LibriSpeech and are comparable.}
  \label{new-speaker-result}
  \centering
  \begin{tabular}{lllll}
    \toprule
    & \textit{new}-eval & test-clean \\
    \midrule
    AGN Conventional & 9.3 dB & x  \\
    AGN Robust    & 9.3 dB & x \\
    Xiao-Attention   & x & 9.8dB   \\ \bottomrule
    \bottomrule
  \end{tabular}
\end{table}

\subsubsection{Stochastic Speaker-Set-Biased - Multiple Speakers in Target and Interferer}
In this experiment there can be multiple speakers in
the target and interferer to demonstrate the strength of AGN against other speaker-biased approaches such as Xiao-Attention. We also make the number of speakers in the mixture stochastic to demonstrate the strength of the speaker-biased approach against to speaker-independent methods such as Deep Clustering (DPCL).

Since $G \in [1, 3]$ and $H \in [1, 3]$ (Eq. \ref{eq:g_sample}) there are between 2 and 6 total speakers in the mixture, and between 1 and 3 speakers to extract. To keep the results comparable to previous work we ensure that there are only two speakers speaking at any given timepoint. This is done by making only one speaker active at any given time point in both the target and interferer signals. Both the target and interferer can be thought of as a conversation between 1 and 3 individuals. In this context the goal is to listen to one conversation and ignore the other. Results are shown in Table \ref{multi-result}, the DPCL result is taken from \cite{isik2016single} An audio sample is available. \footnote{See supplementary audio or \url{https://soundcloud.com/anon-ymous-647326941/sets/multi-attention-example}}.

\begin{table}
  \caption{Pre-Training results for the single-speaker and speaker-set experiments. Only AGN can operate on the stochastic speaker-set task.}
  \label{multi-result}
  \centering
  \begin{tabular}{lllll}
    \toprule
 %   \multicolumn{2}{c}{Part}                   \\
%    \cmidrule(r){1-2}
    & Speaker-Set-Target-Interferer & Single-Target-Interferer \\
    & (\textit{well-known}-eval) & (best result) \\
    \midrule
    AGN      & 9.9 dB & 10.6 dB  \\
    Xiao-Attention & Not Possible &  \hspace{1mm} 9.8 dB \\
    DPCL   & Not Possible & 10.8 dB   \\ \bottomrule
    \bottomrule
  \end{tabular}
\end{table}

\section{Discussion}
In the first pre-training experiment AGN performed 9\% better than the baseline model of \cite{xiao2019single}, which had a comparable architecture. We believe this gain in performance is due to our embedding based approach in contrast to the reference audio approach of \cite{xiao2019single}. Doing so allows for the embedding vector to be a representation built from all utterances of a speaker rather than just a snippet in the reference audio. The disadvantage of the embedding method is that adapting to new speakers requires a second step, fine-tuning.

During fine-tuning we experimented with conventional fine-tuning and robust fine-tuning. Robust fine-tuning gave AGN the ability to extract new speakers without interfering with it's performance on previously learned speakers. Surprisingly, the robust fine-tuning method worked as well as the conventional one, indicating that the neural network has learned an efficient basis for segmenting any speaker, given an optimized speaker embedding. This method did perform 5\% worse than the \cite{xiao2019single} attention model. We believe this is due to overfitting on the well-known speakers during pre-training. Using a meta-learning method like MAML  \citep{finn2017model} might allow AGN to perform better here.

In the final experiment we investigated AGN on the new stochastic speaker-set-biased task. In this task AGN must extract an arbitrary number of speakers given an unknown number of sources. Interestingly, AGN only performed 7\% worse than in the single speaker-biased task. This is surprising because the number of possible speaker-set embeddings is $\sum_{i=1}^{3} \binom{916}{i} = 12809664$. This demonstrates how powerful the simple superposition method is for extracting speakers using only an additive gain factor. Because the number of sources in the mixture was unknown deep clustering (DPCL), nor other speaker-independent methods, could not be applied to this task. Because multiple speakers needed to be extracted at once the method of \cite{xiao2019single} could also not be used.

\section{Conclusion}
Speaker-biased approaches have been of interest recently in the audio deep learning community. However a major limitation of these models when compared to speaker-independent approaches is that they could not extract multiple speakers. This scenario is quite common in our everyday lives, where we need to attend to a small subset of voices present in noisy environments like a restaurant.

Here we developed a method to replicate this human ability using the Attentional Gating Network (AGN). We showed AGN can outperform the speaker-biased approach of \cite{xiao2019single}. In subsequent fine-tuning experiments on \textit{new} speakers, it was shown that a similar performance can be reached by only optimizing the embedding parameters. Finally we demonstrated the most powerful feature of AGN. In the stochastic speaker-set-biased experiment we showed AGN can extract an arbitrary number of voices from a mixture of an unknown number of sources. We summarize the characteristics of speaker-independent, speaker-biased, and speaker-set methods in Table \ref{tb-summary-methods}.

\begin{table}
  \caption{Summary of the advantages and disadvantages between speaker-independent, speaker-biased, and our speaker-set-biased method.}
  \centering
  \begin{tabular}{lllll}
    \toprule
 %   \multicolumn{2}{c}{Part}                   \\
%    \cmidrule(r){1-2}
    & Spk-Independent & Spk-Biased & Spk-Set-Biased \\
    \midrule
    Number of Speakers Separable & All & One & \textbf{All Subsets} \\
    Agnostic to the Number of Speakers & \xmark & \cmark & \cmark \\
    No prior knowledge of Spk required & \cmark & \xmark & \xmark \\

    \bottomrule
  \end{tabular}
  \label{tb-summary-methods}
\end{table}

\subsection{Future Work}
An interesting follow up to the embedding-based approach would be to resolve the remaining disadvantage of speaker-biased approaches, the requirement of prior knowledge. Instead of fine-tuning with a speaker profile of utterances, one could instead learn \textit{new} speakers by fine-tuning the discriminator of a Generative Adversarial Network (GAN)  \citep{goodfellow2014generative}. The discriminator could be trained to identify single speaker speech while AGN could serve as the generator, generating single speaker speech from a mixture of speech. In order to discover new speakers, an additional loss term could be added to push AGN to look for a new embedding that is dissimilar to all embeddings it already knows. In this way AGN could continually learn about new speakers on the fly, with minimal supervision, much like humans do.

\newpage
\bibliographystyle{plainnat}
\bibliography{neurips_2019}

\end{document}